\documentclass[a4paper,11pt]{article}
\usepackage{pos}

\usepackage{color}
\usepackage{tikz}
\usepackage{pgfplots}
\usepackage{pgf-umlsd}
\usepackage[]{fp}
\pgfplotsset{compat=1.17}
\usepackage{hhline} % For hhline (double line) 
\usepackage[percent]{overpic}
\usepackage{rotating}
\usepackage{float}
\usepackage{xspace}
\usepackage{siunitx}
\usepackage{setspace}

\title{Results from the ARIANNA high-energy neutrino detector}
\ShortTitle{Results from the ARIANNA high-energy neutrino detector}

\author[a]{Christian Glaser for the ARIANNA collaboration}

\affiliation[a]{Department of Physics and Astronomy, Uppsala University,\\
SE-75237, Uppsala, Sweden}

\emailAdd{christian.glaser@physics.uu.se}

\abstract{The ARIANNA in-ice radio detector explores the detection of UHE neutrinos with shallow detector stations on the Ross Ice Shelf and the South Pole. Here, we present recent results that lay the foundation for future large-scale experiments. We show a limit on the UHE neutrino flux derived from ARIANNA data, measurements of the more abundant air showers, results from in-situ measurement campaigns, a study of a potential background from internal reflection layers, and give an outlook of future detector improvements.}

\FullConference{%
  9th International Workshop on Acoustic and Radio EeV Neutrino Detection Activities - ARENA2022\\
  7-10 June 2022\\
  Santiago de Compostela, Spain}

\begin{document}
\maketitle

\section{Introduction}
In-ice radio detection is a promising technique to measure the low expected flux of ultra-high-energy neutrinos ($E_\nu >\SI{e17}{eV}$ \cite{Barwick:2022vqt}. It was explored in the smaller test bed arrays ARA and ARIANNA that demonstrated the technique's feasibility using two complementary detector station designs. 
Currently, a larger array of autonomous radio detector stations is being constructed in Greenland (RNO-G) \cite{RNO-G:2020rmc} with about an order of magnitude larger effective volume to neutrinos than previous experiments, which might allow the detection of the first UHE neutrino. At the same time, a much larger in-ice radio array is being planned as part of IceCube-Gen2 \cite{IceCube-Gen2:2021rkf}. Here, we report on how ARIANNA advanced the field of UHE neutrino detection and is paving the way for IceCube-Gen2. 

\section{Versatile + Autonomous Hardware Design}
\begin{figure}[t]
     \centering
     \includegraphics[width=0.6\textwidth]{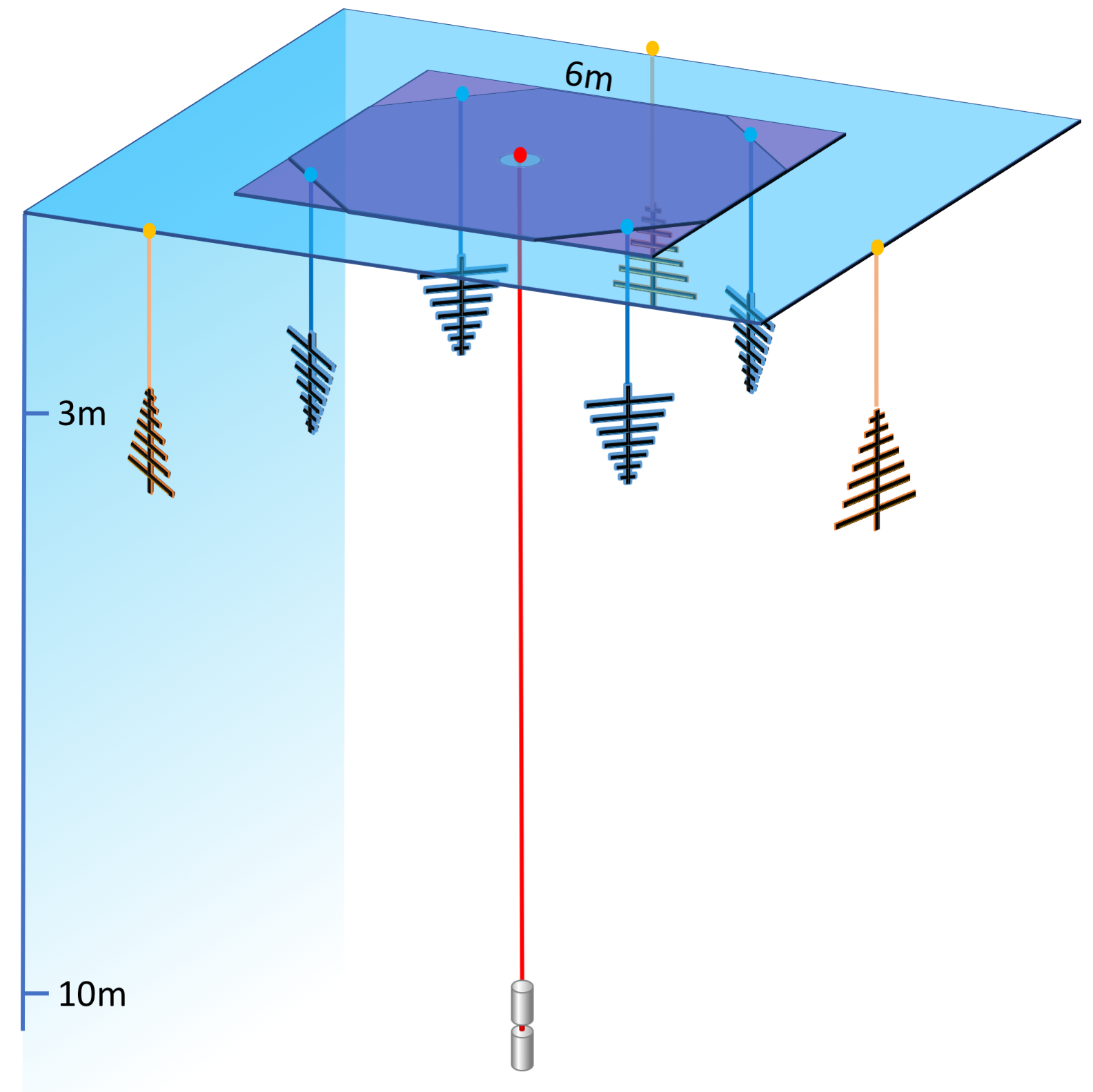}
     \caption{Sketch of the improved shallow station design. The original four downward-facing LPDA antennas are complemented by three upward-facing LPDA antennas for cosmic-ray detection and one dipole to aid background removal and event reconstruction.}
     \label{fig:stationlayout}
 \end{figure}

The ARIANNA concept is based around compact shallow detector stations, i.e., antennas are deployed just underneath the snow surface with a small spatial extend of $\mathcal{O}$(\SI{10}{m}). See Fig.~\ref{fig:stationlayout} for a sketch of the station layout.
The ARIANNA hardware evolved into a versatile Plug'n'Play system that works at essentially any location and has been successfully used on the Ross Ice Shelf \cite{Anker:2019rzo}, the South Pole \cite{ARIANNA:2020zrg} and at Mt. Melbourne in Antarctica \cite{TAROGE:2022soh}. All electronics and batteries are housed in a pelican case with thermal insulation with external ports for the antennas, communication, and power inputs from solar and wind power. This simplifies deployment, as during field installation, only the external ports need to be connected. The station starts up automatically and can be commissioned remotely.

The ARIANNA hardware is based on the SST chip design \citep{Kleinfelder2015}. It currently supports up to 8 input channels on a single board. The input is sampled at \SI{1}{GSPS} or \SI{2}{GSPS} and continuously stored in a switched capacitor array. The SST chip has a precise time synchronization between samples and across channels of less than \SI{5}{ps}, which allows for a precise reconstruction of the signal direction despite the small spatial extent of an ARIANNA station. Following a trigger, 256 samples are digitized with 12 bit ADCs and read into an FPGA and, after that, into an Mbed microprocessor to calculate a second trigger stage and data storage. 

The system has shown to work reliably over several years in Antarctica with only \SI{5}{W} power consumption. Communication is established via the IRIDIUM satellite network or long-range WiFi if available. The station is powered by solar panels and experimentally through wind turbines to provide uptime during the polar winter. A custom wind turbine -- developed and built at Uppsala University -- was successfully tested at Moore's Bay and powered an ARIANNA station for 40\% during the winter \cite{Nelles:2020xtn}. Larger versions of this turbine are under development to increase uptime and provide autonomous power also at the RNO-G site in Greenland and the South Pole, which have more challenging wind conditions. 

\section{The Hexagonal Radio Array - a Test Bed for a Larger Scale Array}
\begin{figure}[t]
    \centering
    \includegraphics[width=0.75\textwidth]{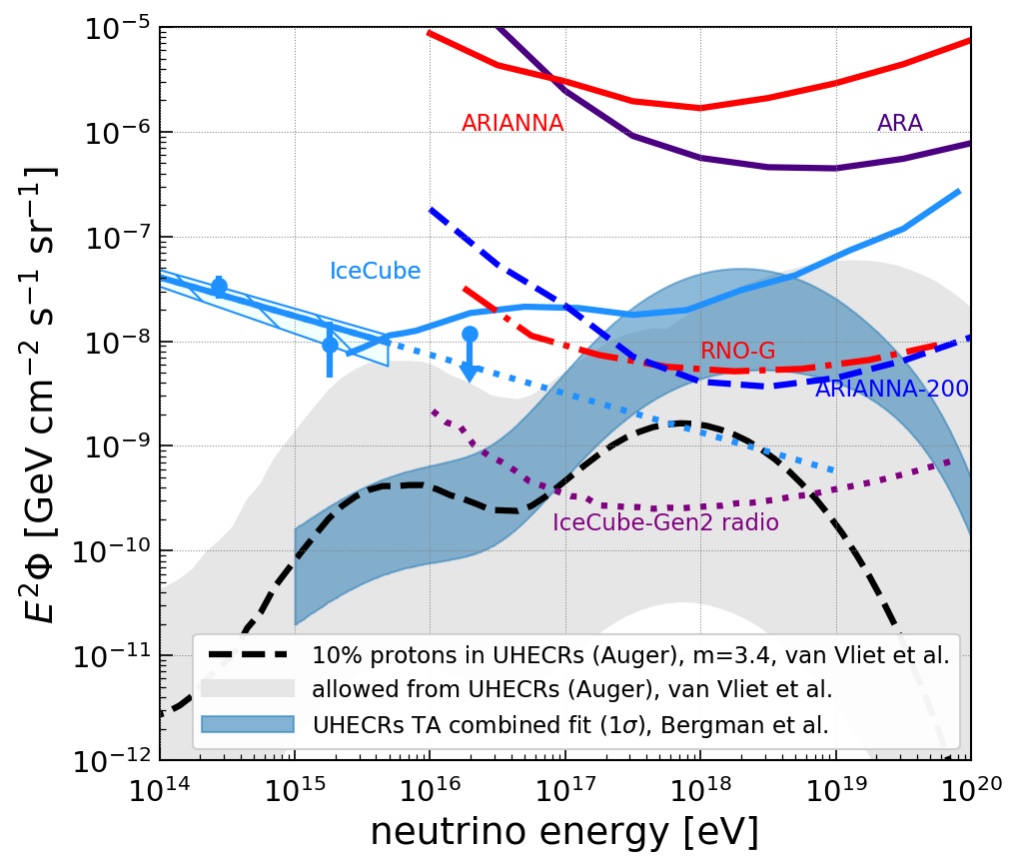}
    \caption{Measured astrophysical neutrino flux by IceCube and limits from different experiments. Also shown is the expected sensitivity of future detectors and predictions of the GZK neutrino flux based on measurements of the cosmic-ray flux. ARIANNA-200 was a concept study for a 200 shallow station array on the Ross Ice Shelf.}
    \label{fig:diffuse}
\end{figure}

The main purpose of ARIANNA was the demonstration of the feasibility of the in-ice radio technique for detecting ultra-high-energy neutrinos. This was done in the \emph{hexagonal radio array}, an array of seven identical radio detector stations, each comprising four downward-pointing LPDA antennas arranged in two orthogonal pairs with a separation of \SI{6}{m} \cite{Barwick:2014rca, ARIANNA:2014fsk}. The array was completed in 2015 and operated successfully until it was decommissioned. ARIANNA demonstrated that an uptime above 90\% is achievable during the summer months when solar power is available \cite{Anker:2019rzo}. A neutrino search was performed using the full data set. A straight-forward template matching analysis yielded no neutrino candidates, with a signal efficiency of 79\%, and a 90\% confidence upper limit on the diffuse neutrino flux we reported of $E^2 \Phi$ = 
\SI{1.7e-6}{GeV cm^{-2} s^{-1} sr^{-1}} 
for a decade-wide logarithmic bin centered at a neutrino energy of \SI{e18}{eV} \cite{Anker:2019rzo}. The differential limit is shown in Fig.~\ref{fig:diffuse}. 
Although not competitive with existing limits from IceCube or the Pierre Auger Observatory, the study showed the general feasibility of in-ice radio neutrino detection.

\section{Improvements - Towards The Next Generation Neutrino Telescope}
Apart from demonstrating the technical feasibility of in-ice neutrino detection, ARIANNA developed several improvements over the original design that will increase the sensitivity to UHE neutrinos and improve the determination of the neutrino's direction, energy, and flavor.

Upward-facing LPDA antennas were added to the station to measure radio flashes from cosmic-ray-induced air showers. The DAQ system was extended accordingly to support eight input channels. This extension serves three purposes: The rejection of radio signals from air showers that can be a critical background in a neutrino search, testing the detector capabilities under realistic conditions (see below in Sec.~\ref{sec:CR}), and enabling cosmic-ray science. To facilitate the latter, a method was developed to determine the cosmic-ray direction and energy from data of a single radio detector station \cite{Welling:2019scz}. 

An approx.~\SI{10}{m} deep dipole antenna was added to the station design, which allows measuring two time-delayed signals from the same in-ice neutrino interaction \cite{Anker:2019rzo}: One direct trajectory to the antenna and one trajectory where the signal is reflected off the surface, called D'n'R (short for Direct and Reflected). This provides a unique signature of in-ice emission and thus can be used to reject anthropogenic background from above the ice. In 2018, an ARIANNA station with these improvements was installed at the South Pole, successfully taking data until the station was decommissioned in Jan. 2021. A neutrino search was conducted using the same template-matching technique as in the previous search of \cite{Anker:2019zcx}. Using only the four downward-facing LPDAs, a similar analysis efficiency was obtained as previously on the Ross Ice Shelf. Using the upward-facing LPDAs and the dipole antenna as additional discriminators led to a further significant improvement in background rejection \cite{leshan_zhao_2022_6785194}. The final station design is shown in Fig.~\ref{fig:stationlayout}. 

The time delay between the direct and reflected signal also allows measuring the distance to the neutrino interaction, which is a crucial parameter for estimating the neutrino energy. In a simulation study, we derived the relation between time delay and distance and studied the corresponding experimental uncertainties in estimating neutrino energies. We found that the resulting contribution to the energy resolution is well below the natural limit set by the unknown inelasticity in the initial neutrino interaction \cite{Anker:2019rzo}. 
Furthermore, if combined with a calibration emitter installed below the ice surface, the D'n'R time delay can be used to track the snow accumulation throughout the year. This was demonstrated on the Ross Ice Shelf and
provided a precision of $\mathcal{O}$(\SI{1}{mm}) in surface elevation, which is much better than that needed to apply the D'n'R technique to neutrinos \cite{Anker:2019rzo}. We further developed this system into an in-situ calibration system also for the index-of-refraction profile by adding a second calibration emitter and determined the optimal positions in a simulation study \cite{Beise:2022stx}.

Another area of improvement is the trigger to increase the sensitivity to neutrinos of each single detector station. First, the trigger bandwidth was optimized \cite{Glaser:2021gps}: The amplitude of the Askaryan signal increases linearly with frequency up to a cutoff frequency that depends on the viewing angle but extends up to \SI{1}{GHz} if the shower is observed close to the Cherenkov angle. However, it was found that the sensitivity to neutrinos can be increased by up to 50\% if the bandwidth is reduced to the lower end of the frequency band of around $80-200~\mathrm{MHz}$. This is a) because the noise RMS decreases with the square root of the bandwidth, b) because antennas are typically more sensitive at low frequencies (the relevant vector effective length is inversely proportional to the frequency and proportional to the square root of the gain), and c) because, for a given amplitude at the detector, the geometry of off-cone events is more favorable where the frequency cutoff quickly drops into the hundred MHz range.

In addition, sensitivity can be achieved by lowering the trigger threshold. However, the challenge is that current Askaryan detectors already operate at such low trigger thresholds so that the vast majority of triggered events are just unavoidable thermal noise fluctuations. The trigger threshold is then set by the maximum data rate the detector can handle. To get around these limitations, we developed a real-time thermal noise rejection algorithm that enables the trigger thresholds to be lowered, which increases the sensitivity to neutrinos by up to a factor of two (depending on energy) compared to the current ARIANNA capabilities \cite{Arianna:2021vcx}. A deep learning discriminator, based on a Convolutional Neural Network (CNN), was implemented to identify and remove thermal events in real-time. The network retains 95 percent of the neutrino signal at a thermal noise rejection factor of \num{e5}. The results of the simulation study were verified in a lab measurement by feeding in generated neutrino-like signal pulses and thermal noise directly into the ARIANNA data acquisition system. More advanced trigger schemes where the antenna signals are processed continuously are under development, as well as a new all-digital data acquisition system. 

Based on these improvements, a concept study of a 200-station array on the Ross Ice Shelf was conducted \cite{Anker:2020lre}, and its expected sensitivity is shown in Fig.~\ref{fig:diffuse}. The improved shallow stations are also an integral part of the design for IceCube-Gen2. 

\section{Simulation and Reconstruction Tools}
The ARIANNA initiative has also given rise to the development of innovative simulation and reconstruction tools. An accurate and adaptable simulation code is of utmost importance for exploring the capabilities of future detectors, as well as for the design and evaluation of reconstruction algorithms. The NuRadioMC simulation code was developed \cite{Glaser:2019cws}, which incorporates our current best knowledge of the physical processes in a flexible way so that different detector designs can be simulated quickly. It was shown that the results agree within a few percent with previous MC codes if the same physics settings were used. NuRadioMC allows for making reliable predictions on the sensitivity of a radio detector and can be used for an estimate of reconstruction performance. The code quickly became the de-facto community standard for in-ice neutrino detection, with a large community of developers continuously improving and enhancing its capabilities (see e.g. \cite{Garcia-Fernandez:2020dhb, Glaser:2021hfi, Oeyen:2021aco, Heyer:2022ttn}). 

At the same time, the reconstruction framework NuRadioReco \cite{Glaser:2019rxw} was developed to facilitate the analysis of experimental data and the development of reconstruction algorithms. Also NuRadioReco was adopted by the community and serves as the basis for an increasing number of studies. In particular, NuRadioMC and NuRadioReco enabled the ARIANNA studies described below. 

\section{Development and Test of Reconstruction Techniques}
The ability to determine the neutrino's direction and energy is an important requirement for in-ice radio detectors. We can't test the reconstruction performance directly by comparing it to an alternative detection technique, as it is done for the radio detection of cosmic rays, because no UHE neutrino has been measured yet, and large future radio arrays expect to measure just a  handful of neutrinos per year at most. However, what can be done is the following: Certain aspects of the reconstruction can be verified experimentally with in-situ tests. This can then be combined with simulation studies to estimate the reconstruction performance for the expected distribution of neutrino signals. 

The reconstruction of the direction and energy of the neutrino requires the measurement of the distance to the neutrino vertex, the signal arrival direction, the viewing angle, and the signal polarization \cite{Barwick:2022vqt}. We experimentally tested the reconstruction of as many of these parameters as possible. Using the residual hole from the South Pole Ice Core Project, radio pulses were emitted from a transmitter located up to \SI{1.7}{km} below the snow surface. By measuring these signals with an ARIANNA surface station, the signal direction and polarization reconstruction abilities were quantified \cite{ARIANNA:2020zrg}. After deconvolving the raw signals for the detector response and attenuation from propagation through the ice, the signal pulses show no significant distortion and agree with a reference measurement of the emitter made in an anechoic chamber. The origin of the transmitted radio pulse was measured with an angular resolution of \SI{0.37}{\degree}, indicating that the neutrino direction can be determined with good precision if the polarization of the radio pulse can be well determined. In the present study, we obtained a resolution of the polarization vector of \SI{2.7}{\degree}. Neither measurement show a significant offset relative to expectation. One limitation of this study was that not the complete phase space of possible trajectories through the ice could be probed, in particular, mostly horizontally propagating signals. More experimental data probing the effects of propagation through ice on the radio signals will be useful, in particular, to probe the effects of birefringence \cite{Jordan:2019bqu,Heyer:2022ttn}.

We developed several complementary reconstruction algorithms and made them available open-source through NuRadioReco. Among these algorithms, the best performing one was the forward folding technique \cite{Glaser:2019rxw}, where an analytical model of the emitted radio pulse is modified to account for propagation effects and compared with the measured signals in all antennas. This allows for the direct determination of the neutrino direction and shower energy. In a Monte Carlo study, we applied this technique to a representative data set of neutrino signals collected in a state-of-the-art shallow detector station equipped with four downward-facing LPDA antennas and one dipole located  at the South Pole (as shown in Fig.~\ref{fig:stationlayout}). Using all triggered events without any quality cuts, we achieved an angular resolution of \SI{3}{\degree} for the neutrino direction for the subset of hadronic showers \cite{GGaswintPhD, ARIANNAICRC2021Direction}. 
Additionally, we developed a deep convolutional neural network to predict the neutrino energy and direction directly from raw data \cite{Glaser:2022lky}. The results of this network are encouraging, with the angular resolution slightly lower than that obtained using the forward folding technique, but therefore able to reconstruct all event topologies, including the more complex $\nu_e$ charged-current interactions.

\label{sec:CR}
Due to the similarities in generated radio signals, cosmic rays are often used as test beams for neutrino detectors. Some ARIANNA detector stations are equipped with antennas capable of detecting air showers. We showed -- using a straightforward template-based search -- that cosmic rays can be identified, and that the detection rate agrees with expectations \cite{Barwick:2016mxm}, which motivates that a similar search strategy would work for neutrinos as well.

Since the radio emission properties of air showers are well understood, and the polarization of the radio signal can be predicted from the arrival direction, cosmic rays can be used as a proxy to assess the reconstruction capabilities of the ARIANNA neutrino detector. We used the data of two stations installed on the Ross Ice Shelf to obtain a sample of cosmic rays and to reconstruct the polarization of cosmic-ray radio pulses \cite{Arianna:2021lnr}. After correcting for differences in hardware, the two stations used in this study showed similar performance in terms of event rate and agreed with simulation. Subselecting high-quality cosmic rays, the polarizations of these cosmic rays were reconstructed with a resolution of \SI{2.5}{\degree} (68\% containment), which agrees with the expected value obtained from simulation. A large fraction of this resolution originates from uncertainties in the predicted polarization because of the contribution of the subdominant Askaryan effect in addition to the dominant geomagnetic emission. Subselecting events with a zenith angle greater than \SI{70}{\degree} removes most influence of the Askaryan emission, and, with limited statistics, we found the polarization uncertainty is reduced to \SI{1.3}{\degree} (68\% containment). 

\section{Outlook}
ARIANNA pioneered and paved the way for UHE neutrino detection. With shallow detector stations on the Ross Ice Shelf and the South Pole, the feasibility of in-ice radio detection was demonstrated, and the capabilities of the detector to measure the neutrino direction and energy were tested in in-situ measurements and using cosmic rays as a test beam. Currently, the Radio Neutrino Observatory in Greenland is under construction. It will be large enough to potentially measure the first UHE neutrino. At the same time, an order of magnitude larger detector is envisioned as part of IceCube-Gen2 at the South Pole (see Fig.~\ref{fig:diffuse}) where the improved shallow station design shown in Fig.~\ref{fig:stationlayout} is an integral part of the design.

\bibliographystyle{ICRC}
\setlength{\bibsep}{2.0pt}
\bibliography{bib}

% \begin{thebibliography}{99}
% \bibitem{...}
% ....

% \end{thebibliography}

\end{document}